\documentclass[10pt,a4paper]{article}
\RequirePackage{amsmath,amssymb}
\RequirePackage[dvipsnames,usenames]{color}

\usepackage{cite}
\usepackage{fullpage}

\usepackage[british]{babel}
\usepackage[latin1]{inputenc}
\usepackage[T1]{fontenc}
\usepackage[final]{showkeys} 
\usepackage[bookmarks]{hyperref}
\usepackage{amsthm}
\usepackage{graphicx}
\usepackage{subfigure}
\usepackage{braket}
\usepackage{mathrsfs}
\usepackage{color}

\usepackage{bm}
\usepackage{amsfonts}
\usepackage{braket}

\newcommand{\bc}{\begin{center}}
\newcommand{\ec}{\end{center}}

\newcommand{\be}{\begin{eqnarray}}
\newcommand{\ee}{\end{eqnarray}}
\renewcommand{\d}{\mbox{${\rm d}$}}

\title{Turnaround size of non-spherical structures}

\author{
Andrea~Giusti\thanks{E-mail: agiusti@ubishops.ca}  
~and   
Valerio~Faraoni\thanks{E-mail: vfaraoni@ubishops.ca}
$\,$
\\
\\
{\em Department of Physics and Astronomy, Bishop's University}
\\
{\em 2600 College Street, Sherbrooke Qu\'ebec, Canada J1M 1Z7}
}

\begin{document}

\maketitle

\begin{abstract}
The turnaround radius of a large structure in an accelerating universe has 
been studied only for spherical structures, while real astronomical 
systems deviate from spherical symmetry. We show that, for small 
deviations from spherical symmetry, the gauge-invariant characterization 
of the turnaround size using the Hawking-Hayward quasi-local mass and 
spherical symmetry still applies, to first order in the cosmological perturbation 
potentials and in the deviations from sphericity. This is the first step to include 
non-spherical systems in the physics of turnaround.
\end{abstract}

\newpage

    \section{Introduction}
\label{sec:1}

The observational discovery that the expansion of the universe currently 
accelerates  \cite{Perlmutter:1997zf, Perlmutter:1998np} constitutes a  
major leap forward in our understanding of the large-scale 
structure of the cosmos. Within the classical picture 
of General Relativity (GR), this phenomenon is accounted for by 
postulating the existence of a mysterious and exotic fluid with negative 
pressure, referred to as dark energy \cite{AmendolaTsujikawa}. In the 
last two decades, several alternatives 
to the {\em ad hoc} dark energy have been proposed, ranging from modified 
theories of 
gravity \cite{Sotiriou:2008rp, DeFelice:2010aj, Nojiri:2010wj, 
Capozziello:2011et, Capozziello:2009nq} to the backreaction of 
inhomogeneities on the large-scale dynamics \cite{Buchert:1999er, 
Rasanen:2011ki, Bolejko:2011jc}.
	
In recent years, the notion of {\em turnaround radius} 
\cite{TR1,TR2,TR3,TR8} has attracted much attention for its potential 
as a tool to test the $\Lambda$ Cold Dark Matter 
($\Lambda$CDM) model of the late time universe 
\cite{TR4,TR5,TR6,TR7}. In an accelerating 
Friedmann-Lema\^{i}tre-Robertson-Walker (FLRW) universe, the 
turnaround radius is the scale at which a spherical shell of test 
particles can no longer collapse because of the accelerated 
cosmic expansion. In other words, if we look at the problem of structure 
formation in a universe filled with 
dark energy, the collapse of {\em local} inhomogeneities is opposed by the 
accelerated expansion of the cosmos, which suggests the existence of a 
critical scale (the turnaround radius) beyond which an overdensity can no 
longer collapse since its self-gravity is overcome by 
the cosmic expansion. With this perspective, we can provide 
a more rigorous definition of turnaround radius based on the competition 
between local and cosmological physics \cite{usJCAP}.  
Working, for simplicity, in the framework of GR and limiting 
ourselves to the first order in the metric perturbations of a spatially 
flat FLRW universe  dominated by a single dark energy fluid (such that $w 
\equiv p/\rho \approx -1$) and restricting to spherically 
symmetric perturbations, the 
Hawking-Hayward quasi-local mass (which reduces to the 
Misner-Sharp-Hernandez mass \cite{Misner:1964je, 
Hernandez:1966zia} in the spherical case) is computed for a spherical 
surface $\tilde{S}$ containing all contributions due to the local 
perturbations. This quasi-local mass can be split into two contributions 
\cite{usPRD}, namely
\be \label{MSH}
\tilde{M} _{\rm MSH} \simeq m \, a + \frac{H^2 \, \tilde{R} ^3}{2} \, ,
\ee
where $a$ is the FLRW scale factor, $H \equiv \dot{a}/a$ is the Hubble 
parameter, $m$ is the Newtonian mass of the perturbation, and $\tilde{R}$ 
is the areal radius of the surface $\tilde{S}$. One 
can 
clearly interpret the first term in the right hand side of Eq.~\eqref{MSH} 
as the contribution due to the energy of the 
local perturbation, whereas the second term is due to the  
cosmological expansion. Hence, a natural definition of turnaround radius 
is obtained by requiring these two terms to be equal \cite{usPRD}. The 
turnaround radius has been discussed also in alternative theories of 
gravity (see {\em e.g.} \cite{TRA1,TRA2,TRA3,TRA4,TRA5,TRA6,TRA7,TRA8}), 
but in this work we focus solely on the standard GR picture.
	
	Thus far, the literature on the turnaround radius has studied only 
spherical structures. The Misner-Sharp-Hernandez mass is defined only in 
the presence of spherical symmetry and is, therefore, inappropriate to 
describe even small 
deviations from sphericity. The Hawking-Hayward mass is a natural 
generalization of the Misner-Sharp-Hernandez one, to which it reduces in 
spherical symmetry, and it is used in this work.

Preliminary observational evidence seemed to  suggest a violation of the  
maximum turnaround radius predicted by the $\Lambda$CDM model 
\cite{Lee:2015upn, Lee:2016oyu, Lee:2017ejv}, which was interpreted as 
preliminary evidence  against the standard model of 
cosmology.  This statement was later  criticized by the same authors 
following further analysis \cite{Lee:2016qpt}, due to the fact that 
realistic celestial structures are not perfectly spherical, as assumed 
instead in theoretical analyses.  This abuse of the spherical 
approximation introduces an error in the 
comparison of theoretical predictions with real-life astronomical systems 
(see Refs.~\cite{Barrow1, Barrow2, Barrow3} for early comments). 
There is little doubt that deviations from spherical symmetry  
introduce departures from the results predicted for spherical shells of 
matter and the present work intends to account for the non-spherical 
features of local overdensities at the theoretical level, in the 
approximation in which these deviations are small. With 
inspiration from recent studies on spheroidal deformations 
of spherical systems \cite{Casadio:2018mry, Rahim:2018nop}, we provide 
the first step in the characterization of the effect of small, 
non-spherical perturbations on the notion of turnaround size in GR.

We follow the notation of Ref.~\cite{Waldbook}. The metric signature is 
$\left( -+++ \right) $, and we use units in which the speed of light $c$ 
and Newton's 
constant $G$ are unity.

\section{Hawking-Hayward mass in the post-Newtonian regime}
\label{sec:2}

In order to discuss the effects of non-spherical perturbations of a FLRW 
geometry on the quasi-local energy, we first discuss the analogous 
situation for post-Newtonian perturbations of a flat background. Then, in 
Sec.~\ref{sec:3}, we transpose our results to a perturbed FLRW space, 
taking advantage of the fact that the FLRW background is conformally flat, 
and of the known transformation properties of the Hawking-Hayward 
quasi-local mass.

The post-Newtonian line element in spherical coordinates $\left( 
\eta, r, \theta, \varphi \right)$ reads 
\be
\d s^2=-\left(1+2\phi_{\rm N} \right) \d \eta^2 + \left( 1-2\phi_{\rm N} 
\right) \left( \d r^2 + r^2 \d \Omega^2_{(2)}  \right) \,, 
\label{postNewtonian}
\ee
where the Newtonian potential is small, $|\phi_{\rm N} |\ll 1$ and 
$\d \Omega_{(2)}^2= \d \theta^2 +\sin^2 \theta \, \d \varphi^2 $ is the 
line element on the unit 2-sphere.

The quasi-local Hawking-Hayward mass \cite{Hawking, Hayward, Hayward2} 
is defined as follows. 
Let $S$ be a spacelike, closed, orientable, 2-surface. Let $\mathcal{R}$ 
be the induced Ricci scalar on $S$ and consider the outgoing and ingoing  
null geodesic congruences emanating from $S$, with expansion scalars 
$\theta_{(\pm)}$ and shear tensors $\sigma_{ab} ^{(\pm)}$, respectively.
Let $\omega^a$ be the projection onto $S$ of the commutator
of the null normal vectors to $S$ (referred to as ``anholonomicity''). 
Let $\mu$ be the volume 2-form of $S$ and $A$  its area. The 
Hawking-Hayward  quasi-local mass is
\be
M_\text{HH} \equiv \frac{1}{8 \pi} \sqrt{\frac{A}{16 \pi}} \int_S \mu 
\left( \mathcal{R} + \theta_{(+)} \theta_{(-)} -\frac{1}{2} \sigma_{ab} 
^{(+)} 
\sigma_{(-)} ^{ab} - 2 \omega_a \omega^a \right) \,. 
\ee
The quasi-local mass for the post-Newtonian geometry~(\ref{postNewtonian}) 
was computed in Ref.~\cite{usPRD}, obtaining 
\be
M_\text{HH} = \frac{1}{8 \pi}  \sqrt{\frac{A}{16 \pi}} \int_S \mu  
\Bigg( \frac{4}{r} \, \frac{\partial \phi_N}{\partial r} +
\frac{2}{r^2} \, \cot \theta \, \frac{\partial \phi_N}{\partial \theta }
+ \frac{2}{r^2}\, \frac{\partial^2 \phi_N}{\partial \theta^2 }
+ \frac{2}{r^2\sin^2 \theta }\, \frac{\partial^2  \phi_N}{\partial 
\varphi^2 } \Bigg)  \label{eq:31}
\ee
to first order in the metric perturbations. This expression is 
gauge-invariant to first order \cite{usPRD}.  

We now want to study small deviations from spherical symmetry, therefore 
we assume that the Newtonian potential of the mass distribution is 
\be
\phi_{\rm N} \left(r, \theta, \varphi \right) = \phi_0 (r)+ \xi \, f \left(r, 
\theta, \varphi \right)  \label{Newtpot}
\ee
where the spherical potential $ \phi_0(r)$ is slightly perturbed, with   
\be 
\mathcal{O}(f) = \mathcal{O} (\phi_0) = \mathcal{O} (\phi_{\rm N}) \, , \qquad  0<\xi \ll 1 
\,.
\ee
We have, therefore, two expansions: a geometric expansion of spacetime 
quantities at the 
post-Newtonian level, and an expansion in small deviations from 
sphericity at the Newtonian level.  We will take the surface $S$ for 
the calculation of the Hawking-Hayward 
mass to be an equipotential surface, $\phi_{\rm N}=$~const., which 
consists of a 
slight modification of a spherical surface $S_0$ of radius $r_0$ (which 
is an 
equipotential surface of the spherical potential $\phi_0$).  In order for 
the perturbed surface $S$ to stay close to $S_0$, one must control the 
perturbation function $f(r, \theta, \varphi)$ so that it does not 
blow up. If this is not done, the function $f$ could grow very fast and 
the modified surface $S$ could be far from $S_0$ even when the expansion 
parameter $\xi $ is small. To avoid such situations, we impose that 
\be
\frac{\partial f}{\partial r} = \mathcal{O} (\xi) \,. \label{mildgradient}
\ee
We can now use the expansion of the Newtonian potential~(\ref{Newtpot}) 
to obtain
\be
\frac{4}{r} \, \frac{\partial \phi_{\rm N}}{\partial r} +
\frac{2}{r^2} \, \cot \theta \, \frac{\partial \phi_{\rm N}}{\partial \theta }
+ \frac{2}{r^2}\, \frac{\partial^2 \phi_{\rm N}}{\partial \theta^2 }
+ \frac{2}{r^2\sin^2 \theta }\, \frac{\partial^2  \phi_{\rm N}}{\partial 
\varphi^2 } 
= 
\nonumber\\
\frac{4}{r} \, \frac{\d \phi_0}{\d r} 
+  \xi \left( 
\frac{2}{r^2} \, \cot \theta \, \frac{\partial f}{\partial \theta}
+  \frac{4}{r} \, \frac{\partial f}{\partial r}
+ \frac{2}{r^2}\, \frac{\partial^2 f}{\partial \theta^2}
+ \frac{2}{r^2\sin^2 \theta }\, \frac{\partial^2 f}{\partial
\varphi^2} \right) 
\ee
to first order in $\xi$ (in the Newtonian expansion) and in $\phi_{\rm N}$ (in 
the post-Newtonian expansion). 

To proceed, we will now show that the surface  of integration $S$ can be  
replaced by the unperturbed surface $S_0$ up to terms of order $\xi^2$.  

Let us begin with a Newtonian argument to show that $S$ can be replaced 
by the 2-sphere $S_0$ in the Newtonian expansion without affecting the 
integral~(\ref{eq:31}) that we want to calculate to first order in the 
parameter $\xi$ and in the metric perturbations $\phi_{\rm N}$.
By perturbing the sphere $S_0$, one adds to the spherical mass density 
$\rho_0(r)$ a density distribution $\delta \rho(r, \theta, \varphi)$, with 
\be
  \rho(r, \theta, \varphi) = \rho_0(r) +\xi\,  \delta \rho(r, \theta, 
\varphi) 
\,.
\ee
Correspondingly, the Poisson equation (which can be obtained as the 
Newtonian approximation of the Einstein equations for the post-Newtonian 
geometry \cite{usPRD})
\be
\triangle \phi = 4 \, \pi \, \rho \, ,
\ee
with $\triangle$ denoting the Laplace operator, splits into the zero and first order equations in $\xi$
\be
\triangle \phi_0 &\!\!=\!\!& 4 \, \pi \, \rho (r) \, ,\\
\triangle f &\!\!=\!\!& 4 \, \pi \, \delta \rho \left( r, \theta, \varphi \right) \,.
\ee
The total Newtonian mass $M$, which coincides with the Hawking-Hayward 
mass $M_{\rm HH}$ to first order, is 
\begin{eqnarray}
M &=& \int_V \d ^3x \, \rho=  \int_V \d^3x \, \rho_0 +\xi  \int_V \d^3x 
\, \delta \rho \nonumber\\
&&\nonumber\\
&=& \frac{1}{4\pi} \left( \int_V \d^3x \, \triangle \phi_0 +\xi  \int_V \d^3x 
\, \triangle f \right)
\nonumber\\
&&\nonumber\\
&=& \frac{1}{4\pi} \left( \int_S \d^2x \, \bm{n} \cdot \bm{\nabla} \phi_0 
 +\xi  \int_S  \d^2 x \, \bm{n} \cdot \bm{\nabla}  f \right) \,, 
\label{last}
\end{eqnarray}
where $V$ is the 3-volume enclosed by the surface $S$ with unit 
normal $\bm{n}=\bm{\nabla}\phi/| \bm{\nabla}\phi | $ (and, 
correspondingly, 
we denote by $V_0$ the 3-volume enclosed by the 2-sphere $S_0$ with unit 
normal $\bm{n}_0 =\bm{\nabla}\phi_0 /| \bm{\nabla}\phi_0 | $), and we 
have used the Gauss theorem in the last line. 
The second integral in Eq.~(\ref{last}), which is weighted by a factor 
$\xi$, can be replaced with an integral over the unperturbed surface 
$S_0$ by introducing only an error of order $\mathcal{O}(\xi^2)$. The first 
integral requires a more detailed discussion.

Let us begin by computing the unit normal $\bm{n}$ to $S$ to first order:
\begin{eqnarray}
\bm{n} &=& \frac{ \bm{\nabla}\phi }{| \bm{\nabla}\phi |} =
\frac{ \bm{\nabla}\phi_0 + \xi \bm{\nabla}f }{  \sqrt{  
\left(   \bm{\nabla}\phi_0 + \xi \bm{\nabla}f\right) \cdot 
\left(   \bm{\nabla}\phi_0 + \xi \bm{\nabla}f\right) }} \nonumber\\
&&\nonumber\\
&\simeq & \frac{ \bm{\nabla}\phi_0 + \xi \bm{\nabla}f }{
|\bm{\nabla}\phi_0 | \sqrt{
1+2\xi  \, \frac{ \bm{\nabla}\phi_0}{ |\bm{\nabla}\phi_0 |} \cdot \frac{ 
\bm{\nabla}f }{|\bm{\nabla}\phi_0 | }}    } \nonumber\\
&&\nonumber\\
&\simeq & \frac{ \bm{\nabla}\phi_0 + \xi \bm{\nabla}f }{
|\bm{\nabla}\phi_0 | \left( 1+\xi \, \bm{n}_0 \cdot \frac{ \bm{\nabla} 
f }{ |\bm{\nabla}\phi_0 | } \right) }\nonumber\\
&&\nonumber\\
&\simeq & \left( \frac{ \bm{\nabla}\phi_0}{|\bm{\nabla}\phi_0 |} +\xi \, 
\frac{ \bm{\nabla}f}{ |\bm{\nabla}\phi_0 |} \right) \left( 1-\xi 
\, \bm{n}_0 \cdot  \frac{ \bm{\nabla}f}{ |\bm{\nabla}\phi_0 | } 
\right)\nonumber\\
&&\nonumber\\
&\simeq & \bm{n}_0  \left( 1-\xi \, \bm{n}_0  \cdot \frac{ 
\bm{\nabla}f}{ |\bm{\nabla}\phi_0 | } \right) + \xi \, \frac{ 
\bm{\nabla}f}{
|\bm{\nabla}\phi_0 | }  \,.
\end{eqnarray} 
The integrand in the first integral of Eq.~(\ref{last}) is then 
\begin{eqnarray}
\bm{n} \cdot \bm{\nabla} \phi_0 & \simeq & 
\bm{n}_0 \cdot \bm{\nabla} \phi_0
+  \xi \left[ - \left( 
\bm{n}_0 \cdot 
\bm{\nabla} f \right) \left( \bm{n}_0 \cdot \frac{ 
\bm{\nabla}\phi_0 }{| \bm{\nabla}\phi_0 |} \right) 
+
\frac{ \bm{\nabla}f }{| \bm{\nabla}\phi_0 |} \cdot  \bm{\nabla}\phi_0 
\right] \nonumber\\
&&\nonumber\\
&=& \bm{n}_0 \cdot \bm{\nabla} \phi_0 + \xi \left[  - |\bm{n}_0|^2  
\left( \bm{n}_0 \cdot \bm{\nabla}f \right) + \bm{\nabla}f \cdot 
\bm{n}_0 \right] \nonumber\\
&&\nonumber\\
&=& \bm{n}_0 \cdot \bm{\nabla} \phi_0 +\mathcal{O}(\xi^2) \,.  
\label{urca}
\end{eqnarray}
The equation of the unperturbed surface $S_0$ (an equipotential surface 
of the unperturbed potential) is $\phi_0(r)=
\mbox{const}$, or $r=r_0$. The surface $S$ is an equipotential surface of 
the perturbed potential 
\be
\phi_0(r)+\xi \, f(r, \theta, \varphi)= \mbox{const.} \label{czz} 
\ee
The  points of $S$ are located at radius $r=r_0 +\delta_1 r$, with 
$\delta_1 
r/r_0 =\mathcal{O}(\xi)$. Therefore, the equation~(\ref{czz}) describing $S$ 
can be written as
\be
\phi_0 (r_0+\delta_1 r)+\xi \, f( r_0 + \delta_1 r, \theta, \varphi) =
\mbox{const.}
\ee
and expanded to first order in $\xi$ as 
\begin{eqnarray}
\phi_0 (r_0) + \frac{\d \phi_0}{\d r} \bigg|_{r_0}\delta_1 r + \ldots +
\xi \, f( r_0, \theta, \varphi) +\xi \, \frac{\partial f}{\partial r }\bigg|_{( 
r_0, \theta, \varphi)} \delta_1 r + \ldots = \mbox{const.} \quad
\end{eqnarray}
Since $\phi_0(r_0)$ and $ \d \phi_0/\d r |_{r_0}$ are constants, 
Eq.~\eqref{mildgradient} yields, to first order, 
\be 
f( r_0, \theta, \varphi) = \mbox{const.}
\ee
In other words, as for $S_0$, the surface $S$ can be described by letting 
only the polar coordinates $\theta$ and $\varphi$ vary. This means that, 
to 
first order in $\xi$, the integrals over $S$ in Eq.~(\ref{last}) can be  
replaced by integrals over the unperturbed surface $S_0$. By combining 
this result with Eq.~(\ref{urca}), Eq.~(\ref{last}) becomes
\be
4\pi M = \int_{S_0} \d^2 x \, \bm{n}_0 \cdot \bm{\nabla}\phi_0 + \xi 
\int_{S_0} \d^2 x \, \bm{n}_0 \cdot \bm{\nabla} f \,. \label{urca2}
\ee 
We now use the assumption~(\ref{mildgradient})  that $f$ varies gently in 
the radial direction, whcih implies that the second integral in the right 
hand side of  
Eq.~(\ref{urca2}) is of order $\mathcal{O} (\xi^2$) and can be dropped. 
(For special perturbations in which the 
perturbation potential does not depend on the radial coordinate, 
$f=f(\theta, \varphi)$, this integral is identically zero since 
$\bm{n}_0 \cdot \bm{\nabla} f= \d f /\d r=0$.)  As a result, 
\be
M_{\rm HH} =\frac{1}{4\pi} \int_{S_0} \d^2 x \, \bm{n}_0 \cdot 
\bm{\nabla}\phi_0 
+\mathcal{O}(\xi^2) \equiv M_0 +\mathcal{O} (\xi^2) \,,
\ee
where $M_0$ is the mass associated with the unperturbed spherical 
configuration. One can make sense of this result by realizing that the 
total mass $M=M_0+\delta M$ is obtained by adding  to the mass $M_0$ an 
additional mass $\delta M$ of order 
\be
 \delta M \sim  \delta V \, \xi \, \delta \rho \sim (4 \, \pi \, r_0^2 \, \delta_1 r) 
\,\xi \, \delta \rho =\mathcal{O} (\xi^2) \,.
\ee

\section{Hawking-Hayward mass in a perturbed FLRW cosmos}
\label{sec:3}
  
We now turn our attention to the problem of the turnaround radius in 
cosmology. We assume an accelerated FLRW background universe perturbed by 
inhomogeneities described by metric perturbations $\phi_{\rm N}$ of spatial 
extent much smaller than the Hubble radius $H^{-1}$.  It is convenient to adopt the 
conformal Newtonian gauge in which the line element assumes the form
\be
\d \tilde{s}^2=a^2(\eta) \left[ -\left(1+\phi_{\rm N} \right) \d \eta^2 + 
\left(1-\phi_{\rm N} \right)\left( \d r^2 +r^2 \d \Omega_{(2)}^2 \right)\right] \,,
\label{postFLRW}
\ee
where $\phi_{\rm N}(\bm{x})$ describes the perturbation and $\eta$ denotes the 
conformal time of the FLRW background. In this gauge, the line element is 
explicitly conformal to the post-Newtonian geometry~(\ref{postNewtonian}). 
The Hawking-Hayward mass $ \widetilde{M}_{HH}$ contained in a  topological 
2-sphere $S$  in the spacetime~(\ref{postFLRW}) was computed in 
Ref.~\cite{usPRD} in two ways. The first method employs the transformation 
property of the Hawking-Hayward quasi-local mass under the conformal 
transformation of the spacetime metric $g_{ab} \rightarrow 
\tilde{g}_{ab} =\Omega^2 g_{ab}$. The result is \cite{usCQG}
\be
M_{\rm HH} \rightarrow \widetilde{M}_{\rm HH} = \sqrt{ 
\frac{\tilde{A}}{A}}\, M_{\rm HH} 
+\frac{1}{4\pi} \sqrt{ \frac{\tilde{A}}{4\pi}}\, \int_S \mu \left[ h^{ab} 
\left( \frac{ 2\nabla_a \Omega \nabla_b \Omega}{\Omega^2} -\frac{\nabla_a 
\nabla_b \Omega}{\Omega} \right) -\frac{ \nabla^c \Omega \nabla_c 
\Omega}{\Omega^2} \right] \,,
\ee
where $h_{ab}$ is the 2-metric induced on the 2-surface $S$ by the 
post-Newtonian metric $g_{ab}$ of Eq.~(\ref{postNewtonian}) and $\nabla_a$ 
is the covariant derivative of $g_{ab}$. The conformal factor 
$\Omega=a(\eta)$ depends only on the conformal time $\eta$ of the 
FLRW background.  Assuming that the 2-surface $S$ is a small deformation 
of an  
unperturbed 2-sphere $S_0$ of the 
Minkowski background of $g_{ab}$ (in which $\phi_{\rm N}$ is a spherical  
potential $\phi_0(r)$ plus a non-spherical deviation $\xi \, f(r, \theta, 
\varphi)$ parametrized by $\xi$) with radius $R$, the conformal 
transformation maps $S_0$ into a comoving 2-sphere $\widetilde{S}_0$  of 
areal radius $\widetilde{R}=a \, R$, while $\widetilde{S}$ ({\em i.e.}, 
the conformal cousin 
of $S$) is obtained by introducing small non-spherical deformations of 
$\widetilde{S}_0$ 
parametrized by $\xi$. The Hawking-Hayward mass contained in $\widetilde{S}$ 
is \cite{usPRD}
\be
\widetilde{M}_{\rm HH}= \Omega \, M_{\rm HH} +\frac{R \, \Omega_{, 
\eta}}{4 \pi} \left(  \int_{\tilde{S}_0} \mu \, {\phi _{\rm N}}_{, \eta} - 
\frac{ \Omega_{,\eta}}{\Omega} \int_{  \tilde{S}_0} \mu \, \phi_{\rm N} 
\right) + \frac{R^3 \, \Omega^2_{,\eta}}{2\Omega} 
\label{Delta}
\ee
to first order in the perturbations $\phi_{\rm N}$. 
Since, to first order in $\phi_{\rm N}$, the Hawking-Hayward mass is 
gauge-independent, this result is manifestly gauge-invariant to this order 
although it is derived using the conformal Newtonian 
gauge~(\ref{postFLRW}).\footnote{See also the discussion in 
Ref.~\cite{usPRD}.}  The mass \eqref{Delta} 
can also be calculated directly from the post-FLRW 
metric~(\ref{postFLRW}), obtaining the same result \cite{usPRD}.

\section{Turnaround size of a large structure}
\label{sec:4}

If the metric perturbation $\phi_{\rm N}$ is spherically symmetric, 
$\phi_{\rm N}=\phi_0(r)$, then the Hawking-Hayward mass reduces to  
\cite{usPRD, usJCAP}
\be
\widetilde{M}_{\rm HH}= m \, a+ \frac{H^2 \, \widetilde{R}^3}{2} 
\left(1-\phi_{\rm N} \right)  \simeq m \, a+ \frac{H^2 \, 
\widetilde{R}^3}{2}  \,,\label{ancora}
\ee
where $m$ is the Newtonian mass computed from the post-Newtonian line 
element~(\ref{postNewtonian}) and $\widetilde{R}=a(\eta) \, R$ is the areal radius 
of the sphere $\widetilde{S}_0$. In the last equality on the right hand side 
of Eq.~(\ref{ancora}), the correction $-\phi_{\rm N} \, \widetilde{R} \, (H 
\tilde{R})^2$ is neglected for structures of size $\widetilde{R}$ much 
smaller than the Hubble radius (this term is of second order in 
$\widetilde{R}/H^{-1}$).

The current literature on the turnaround radius is restricted to this 
spherical situation \cite{TR1,TR2,TR3,TR8, 
TRA1,TRA2,TRA3,TRA4,TRA5,TRA6,TRA7,TRA8}. In this context, the turnaround 
radius is obtained when the ``local'' and the ``cosmological'' 
contributions to the right hand side of Eq.~(\ref{ancora}) are equal 
\cite{usJCAP}, which yields 
\be 
\widetilde{R}_c= \left( \frac{2ma}{H^2} 
\right)^{1/3} \,. 
\ee 
Other definitions of turnaround radius exist 
in the literature. For example in Ref.~\cite{TRA1}, which considers 
scalar-tensor gravity, the author considers timelike radial geodesics and 
sets the radial acceleration to zero to define the turnaround radius. In 
the case of scalar-tensor gravity considered in that reference, there is 
no accepted definition of Hawking-Hayward mass, which therefore cannot be 
used. In our article, restricted (at least for the moment) to GR, we 
prefer to define the turnaround radius using the Hawking-Hayward mass and 
its splitting in two parts. This definition has the advantage of being 
covariant and gauge-invariant (the same cannot be said, at least a priori, 
for the alternative definition of Ref.~\cite{TRA1}).

In an accelerated FLRW universe dominated by dark energy with equation of 
state $P_{\rm DE}=w \, \rho_{\rm DE}$, the (unperturbed) Friedmann equation 
\be
H^2= \frac{8\pi}{3} \, \rho_{\rm DE} 
\ee
where $\rho_{\rm DE} = \rho_0 \, a^{-3(1+w)}$ for $w \neq -1$,  gives 
$a(t) = a_\ast \, t^{ \frac{2}{3(1+w)}}$ in terms of the {\em comoving } 
time $t$ of the 
background (related to $\eta$ by $\d t=a \, \d \eta$). These relations 
then yield
\be
\widetilde{R}_{\rm c} (a) =  \left( \frac{3m}{4\pi \rho_0} \, a^{3w+4} \right)^{1/3}  
\equiv \widetilde{R}_{\rm c} ^{(0)} \, a^{w+4/3}\,.
\ee
In terms of the redshift factor $z= \frac{a_0}{a}-1$, one has
\be
\widetilde{R}_{\rm c} (z) =  \frac{\widetilde{R}_{\rm c} ^{(0)} \, a_0 ^{w + 4/3}}{(1+z)^{w+4/3}} \,.
\ee
If the turnaround radius could be determined accurately by astronomical 
observations, it would be possible to obtain information about the 
equation of state parameter of dark energy as a function of redshift
\be \label{w_z}
w (z) = -\frac{4}{3} + 
\frac{\ln \left( \widetilde{R}_{\rm c} ^{(0)} \, a_0 ^{w + 4/3} / \widetilde{R}_{\rm c} \right)}{\ln(1+z)} \,. 
\ee
The latter could then be used to constrain $w$ if $m \, a$ and 
$\widetilde{R}_{\rm c}$ were known.

Let us consider now more realistic situations in which matter is not 
distributed spherically and the metric perturbation is described by 
$\phi_{\rm N} (\bm{x})=\phi_0(r)+\xi \, f(r, \theta, \varphi)$ with $\xi \ll 1$ in 
Eq.~(\ref{postFLRW}). Correspondingly, the 2-sphere $S_0$ is perturbed and 
its conformal cousin $\widetilde{S}_0$ in the perturbed FLRW 
space~(\ref{postFLRW}) is slightly deformed. However, to first order in 
the parameter $\xi$, $ M_{\rm HH}$ does not change. In the right hand side of 
Eq.~(\ref{Delta}) it is ${\phi _{\rm N}}_{, \eta}=0$ and the second term 
\be
\frac{R \, \Omega^2_{,\eta}}{4\pi \Omega} \int_{ \tilde{S}_0} \mu \phi_N 
\approx \frac{ \dot{a}^2 \, a \, R}{4\pi} \, 4\pi R^2 \phi_{\rm N} \approx 
\left(H \widetilde{R} \right)^2 \widetilde{R} \,\phi_{\rm N} 
\ee
is of second order in $\widetilde{R}/H^{-1}$ and, therefore, negligible for 
structures of size $\widetilde{R}$ much smaller than  the Hubble 
radius $H^{-1}$, such as those observed in practice in attempts to 
determine the critical radius \cite{Lee:2015upn, Lee:2016oyu, 
Lee:2017ejv, Lee:2016qpt}.  As a result, the Hawking mass 
\be
\widetilde{M}_{\rm HH} \simeq m \, a + \frac{H^2 \, \widetilde{R}^3}{2}  \label{DeltaDelta}
\ee
corresponding to the metric perturbation $\phi_{\rm N}=\phi_0(r) +\xi \, f(r, 
\theta, \varphi)$ coincides, to order $\mathcal{O}(\xi)$, with the 
Hawking-Hayward mass corresponding to $\phi_0(r)$. As a consequence, the 
definition of turnaround radius obtained by equating the two terms on the 
right hand side of Eq.~\eqref{DeltaDelta} is insensitive to the change 
$\phi_0(r) \rightarrow \phi_0(r)+\xi \, f(r, \theta, \varphi)$, to first 
order in $\xi$. Therefore, when one examines a cosmic structure that 
deviates from sphericity due to the $\xi$-corrections in the metric 
perturbation $\phi_{\rm N}=\phi_0(r)+\xi \, f(r, \theta, \varphi)$, the changes  
to the definition and the value of the turnaround radius are negligible 
(to first order in $\xi$). This result constitutes a first step toward 
obtaining cosmological information using astronomical determinations of 
the turnaround size of cosmic structures which, realistically, depart from 
perfect sphericity.

\section{Conclusions}
\label{sec:5}

There is by now little doubt that the use of theoretical predictions for 
the turnaround size of cosmic structures made under the assumption that 
the latter are spherical introduces a significant error when comparing 
theory and observations. All theoretical studies of the turnaround radius 
thus far have assumed spherical symmetry, while celestial systems are not 
spherical. Current attempts to pin down the turnaround size with 
astronomical observations would be helped considerably if the theory took 
into account deviations from spherical symmetry. As a first step in this 
direction, we have developed the idea that, given the size of the error 
that is deemed acceptable from the observational point of view, deviations 
from spherical symmetry can be neglected in the estimate of the turnaround 
size only if the post-Friedmannian potential describing the local 
inhomogeneity in the metric~(\ref{postFLRW}) deviates from spherical by 
less than this error (quantified by our parameter $\xi$). 

When larger deviations are allowed, second and higher order effects in the 
expansion in $\xi$ must be taken into account in both the integrand 
appearing in the integral of Eq.~(\ref{Delta}) and in the deformation of 
the equipotential surface $\widetilde{S}$ used to compute the quasi-local 
mass. This more refined analysis of the Hawking-Hayward mass and the 
necessary tools will be developed elsewhere.

\section*{Acknowledgments}

This work is supported by Bishop's University and by the Natural Sciences 
and Engineering Research Council of Canada (Grant No.~2016-03803 to V.F.).

\end{document}